\documentclass[twocolumn,superscriptaddress]{revtex4-2}
\usepackage{xcolor}
\usepackage{graphicx}
\usepackage{dcolumn}
\usepackage{bm}
\usepackage[colorlinks=true, a4paper=true, pdfstartview=FitV,
linkcolor=blue, citecolor=blue, urlcolor=blue]{hyperref}

\newcommand{\bea}{\begin{eqnarray}}
\newcommand{\eea}{\end{eqnarray}}
\begin{document}
\definecolor{mygreen}{HTML}{006E28}
\newcommand{\ped}[1]{{\color{mygreen}\textbf{?PED:}  #1}}
\newcommand{\trom}[1]{{\color{red}\textbf{?TR:} \color{red} #1}}
\newcommand{\ya}[1]{\textbf{?YS:} {\color{blue} #1}}
\newcommand{\aw}[1]{{\color{magenta}\textbf{?AW:}  #1}}

\title{Oscillons in gapless theories}
\author{P. Dorey}

\affiliation{Department of Mathematical Sciences, Durham University, Durham DH1 3LE, UK}

\author{T. Roma\'{n}czukiewicz}
\affiliation{
 Institute of Theoretical Physics,  Jagiellonian University, Lojasiewicza 11, 30-348 Krak\'{o}w, Poland
}

\author{Y. Shnir}%

\affiliation{
 Institute of Physics, Carl von Ossietzky University of Oldenburg, Oldenburg D-26111, Germany}
\affiliation{Hanse-Wissenschaftskolleg, Lehmkuhlenbusch 4, 27733 Delmenhorst, Germany
}

\author{A. Wereszczy\'{n}ski}

\affiliation{
 Institute of Theoretical Physics,  Jagiellonian University, Lojasiewicza 11, 30-348 Krak\'{o}w, Poland
}
 \affiliation{Department of Applied Mathematics, University of Salamanca, Casas del Parque 2 and Institute of Fundamental Physics and Mathematics,
University of Salamanca, Plaza de la Merced 1, 37008 - Salamanca, Spain
}

\affiliation{International Institute for Sustainability with Knotted Chiral Meta Matter (WPI-SKCM2), Hiroshima University, Higashi-Hiroshima, Hiroshima 739-8526, Japan}

\begin{abstract}
We show that large scale oscillons, i.e., quasi-periodic, long living particle like solutions, may exist in massless theories, too. Their existence is explained using an effective (smeared) mass threshold which takes into account nonlinear (finite) perturbations.  
\end{abstract}

\maketitle

\section{\label{sec:motiv}Motivation}
Oscillons \cite{G} are quasi-periodic nonperturbative localized excitations with exceptionally long lifetimes \cite{BM, GS}. They are found in many models and in various dimensions, and have applications extending from phase transitions in condensed matter to astrophysics and cosmology \cite{CGM, G-cosm, A, LT}. However, they are still quite mysterious, and their surprising stability has yet to be fully understood. In particular, it is not related to any conservation law nor it is based on any topological reasons, although a picture of adiabatic invariants does shed some light on the question \cite{K, KT}. Recently a close relation between oscillons and sphalerons has been suggested \cite{MR} (further studied in \cite{OQRW}), which underlines their importance in the dynamics of nonlinear systems.  

Being exceptionally long-lived, and performing a huge number of oscillations, an oscillon changes its amplitude, width and fundamental frequency $\omega$ very slowly, losing energy by emission of radiation. Importantly, the fundamental frequency is {\it always below} the mass threshold $m$ of the theory, $\omega < m$, where $m$ is the mass of 
infinitesimally small (meson) perturbations above the vacuum.  
Therefore, oscillons can only radiate due to nonlinear effects through  second \cite{Salmi:2012ta} (or even higher \cite{DRS}) harmonics.
This  prevents the rapid dissipation of energy stored in the oscillon.
In fact, it is commonly assumed that existence of a mass gap in the linear perturbation of the field, i.e., a non-zero quadratic term of the potential in the vicinity of the vacuum, is a {\it necessary condition} for an appearance of oscilllons \cite{A, Amin-1}. 

In the present work we demonstrate that this condition can be relaxed. Concretely, we show that large amplitude oscillons can easily exist in gapless theories. This enlarges the range of application of oscillons to include certain massless models, rendering them more common, and therefore more important, than previously thought. Unlike  more--usual oscillons,
these oscillons radiate through their first harmonics. Nonetheless, they can live for a long time, performing a large number of oscillations. Furthermore, in higher dimensions their lifetimes can sometimes be significantly longer than those of oscillons in a similar but massive theory.

\section{\label{sec:kink} Large oscillons in massless models}
\begin{figure}
\includegraphics[width=1.0\columnwidth]{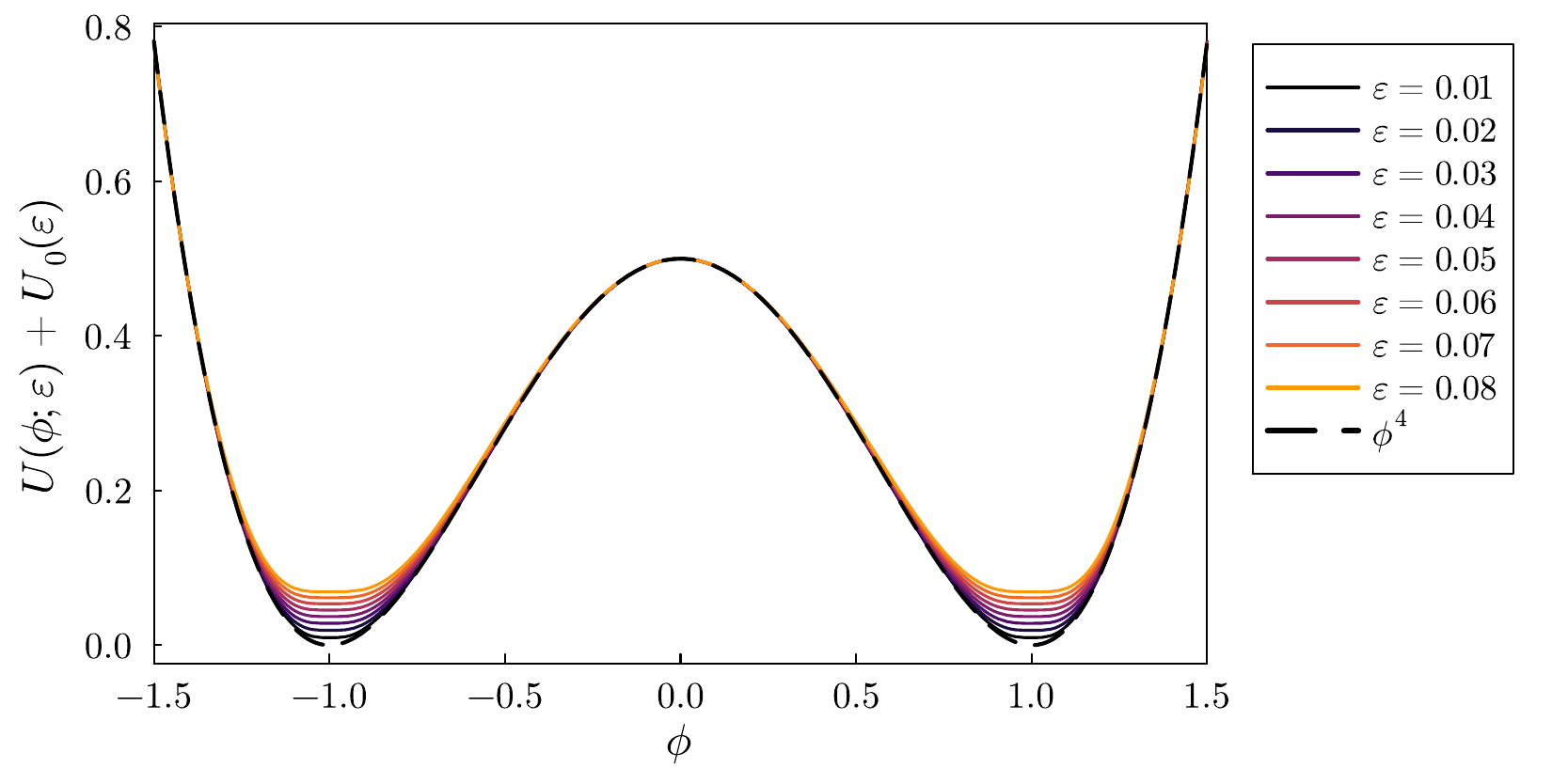}
 \caption{The massless potential (\ref{potential}) for various $\epsilon$.} 
 \label{pot-plot}
 \end{figure}

Let us consider a real scalar field theory in $(d+1)$ dimensions
\begin{equation}
L= \int \left( \frac{1}{2}\phi^2_t- \frac{1}{2} (\nabla \phi )^2 - U(\phi)\right) d^dx. \label{model}
\end{equation}
In our work we assume $d=1$ or 3. 
Next, we choose the potential $U(\phi)$ to have a form very similar to the simplest, most prototypical, double vacuum potential, i.e.,  $\phi^4$,  except for the near vacuum regime. In this region, we assume that instead of a quadratic approach, which provides a nonzero mass of infinitesimally small perturbations, it has a {\it quartic} behaviour. These requirements are fulfilled by the following particular choice of the potential  
\begin{equation}
U(\phi)=\frac{W^2}{W+\epsilon}, \hspace*{0.5cm} W=\frac{1}{2} (1-\phi^2)^2,  \label{potential}
\end{equation}
where $\epsilon$ is a small parameter. This is plotted in Fig.~\ref{pot-plot} where, in addition, we shifted $U$ by a constant $U_0(\epsilon)=-1/(4\epsilon+2)+1/2$ to visualize the near vacuum regime better. Obviously, increasing $\epsilon$ makes the potential flatter in the vicinity of the vacua. 

Importantly, due to the fact that the potential is not quadratic at the vacuum
$U \approx \frac{1}{4\epsilon} (1-\phi)^4(1+\phi)^4$,
the mass of (infinitesimally) small perturbations vanishes, 
\begin{equation}
\left. m^2=\frac{d^2 U}{d\phi^2} \right|_{\phi=\pm 1}=0. \label{mt}
\end{equation}
Thus, as required, there is no mass gap in this model. Therefore, according to the current state of art there should be no oscillon in this theory. In particular, the usual perturbative expansion \cite{Fod} does not find any oscillons.

\begin{figure}
\includegraphics[width=1.0\columnwidth]{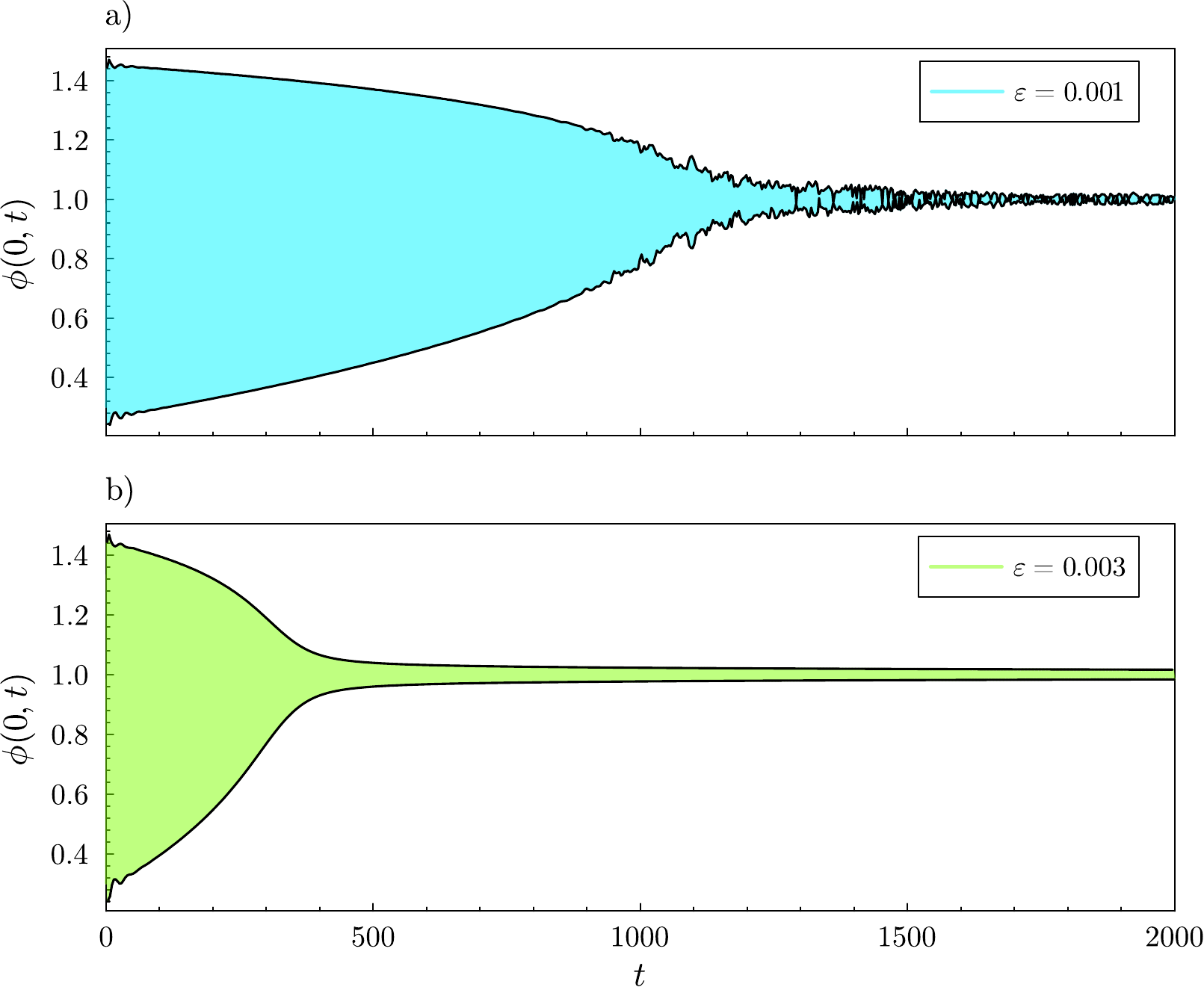}
 \caption{Dynamics of  oscillons in the massless deformation of the $\phi^4$ model in (1+1) dimensions. Value of the field at the origin for $\epsilon=0.002$ and $0.003$.} 
 \label{osc-plot}
 \end{figure}
\begin{figure}
\includegraphics[width=1.0\columnwidth]{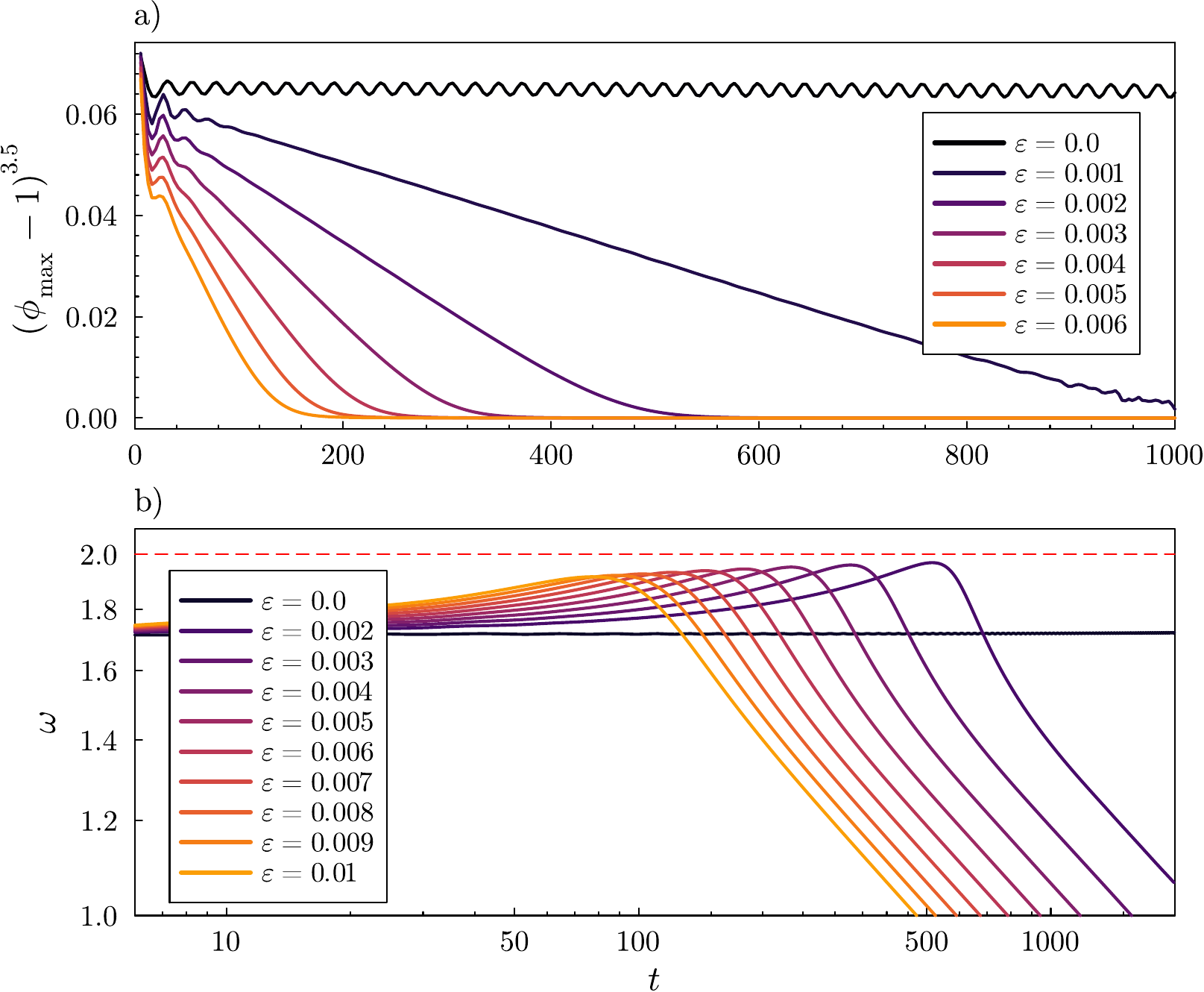}
 \caption{Dynamics of  oscillons in the massless deformation of the $\phi^4$ model in (1+1) dimensions. {\it Upper:} Decay of the amplitude. {\it Lower:} Evolution of the frequency of oscillons.} 
 \label{amplitud-plot}
\end{figure}

 To verify the existence of long-lived excitations
we start in one spatial dimension and use Gaussian  initial data: $\phi(x,0) = 1-0.7e^{-0.204x^2}$ and $\partial_t \phi(x,0)=0$. In Fig.~\ref{osc-plot} we show the time dependence of the field at the origin for $\epsilon=0.002$ and 0.003. As is clearly visible, the field oscillates with a significant amplitude for a large number of oscillations, extending from e.g., $\sim 300$  for $\epsilon=0.003$ to much greater numbers as $\epsilon$ decreases. In fact, the lifetime of the oscillon in (1+1) grows as $\epsilon$ tends to zero and we approach the $\phi^4$ model. Although the lifetime of oscillons in these massless deformed models is shorter than that of a standard oscillon in one space dimension - see Fig.\ \ref{amplitud-plot} - we will see later that this distinction goes away in higher dimensions, justifying our calling these excitations oscillons.

 It is also clearly seen that the amplitude of the oscillations slowly decreases with time.
 As expected this happens due to a direct, but small, coupling of the oscillon's quasi-periodic motion to radiation. In other words, there is a continuous but unexpectedly small leakage of the energy of the oscillon to radiation due to the lack of the mass threshold. Interestingly, we found a universal law governing the amplitude decay in this regime. Namely,
 \begin{equation}
     \phi_{max}=1+c(t_0-t)^\alpha,
 \end{equation}
 where $\alpha \approx 0.3-0.4$ and $c,t_0$ are $\epsilon$ (model) depending constants, see Fig.\ \ref{amplitud-plot}, upper panel. 

During the slow decay  the fundamental frequency of the oscillations grows linearly with time from the value characteristic for oscillons in the usual $\phi^4$ model up to $\omega_{crit}\approx 2$, see Fig.\ \ref{amplitud-plot}, lower panel. At this point the oscillon approaches the minimal, critical amplitude $\phi_{crit}$ and, after this point, it rapidly ceases to exist, quickly radiating all the remaining energy. Thus, there are no oscillons with amplitude smaller than the critical one. 

One immediately recognizes that the critical frequency is very close to the value of the mass threshold in the $\phi^4$ theory. Of course, in our models there is no mass gap in the {\it linear} perturbation analysis. Thus, the appearance of this value must be of a nonlinear nature. 

\begin{figure}
\includegraphics[width=1.0\columnwidth]{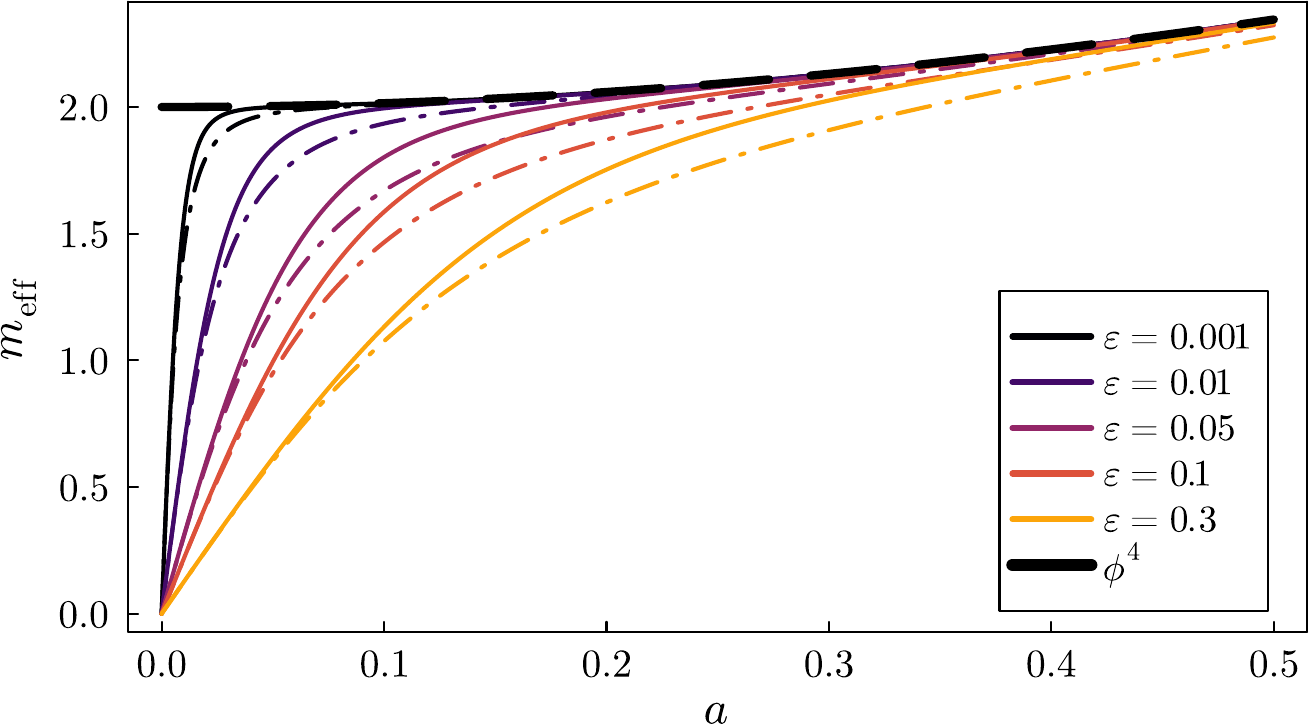}
 \caption{The smeared (finite perturbation) mass threshold: step function (full line) and Gaussian smearing (dashed-dotted line).} 
 \label{mass-plot}
 \end{figure}

To  understand this phenomenon qualitatively we define an {\it effective mass threshold}, which in contrast to the usual mass threshold (\ref{mt}), takes into account finite perturbations, too. In other words, this quantity is determined not only by $U''$ at single point, i.e., the vacuum but rather by its behaviour at the vicinity of the vacuum. Specifically,
\begin{equation}
m^2_{\textrm{eff}}(a)=  \int_{-\infty}^{\infty} \frac{w((\phi-1)/a)}{a} \frac{d^2 U}{d\phi^2} d\phi,  
\end{equation}
where $w(\phi)$ is a symmetric weight function, normalized to unit area and second moment. Here $a$ can be thought as the size of perturbations near to the center of the oscillon (the field amplitude). The usual mass threshold is obtained in the $a \to 0$ limit. In Fig.\ \ref{mass-plot} we present the smeared mass threshold for different $\epsilon$ assuming the step-function smearing $w(\phi)=\theta(1-|\phi|)/2$. Importantly, $m_{\text{eff}}(a)$ is weakly affected by other choices of the weight functions as e.g., Gaussian weight $w(\phi)=\exp(-\phi^2)/\sqrt{\pi}$. Nonzero value of $m_\textrm{eff}$ explains why the oscillons do exist although the perturbative mass threshold vanishes. 

\begin{figure}
   \includegraphics[width=1.0\columnwidth]{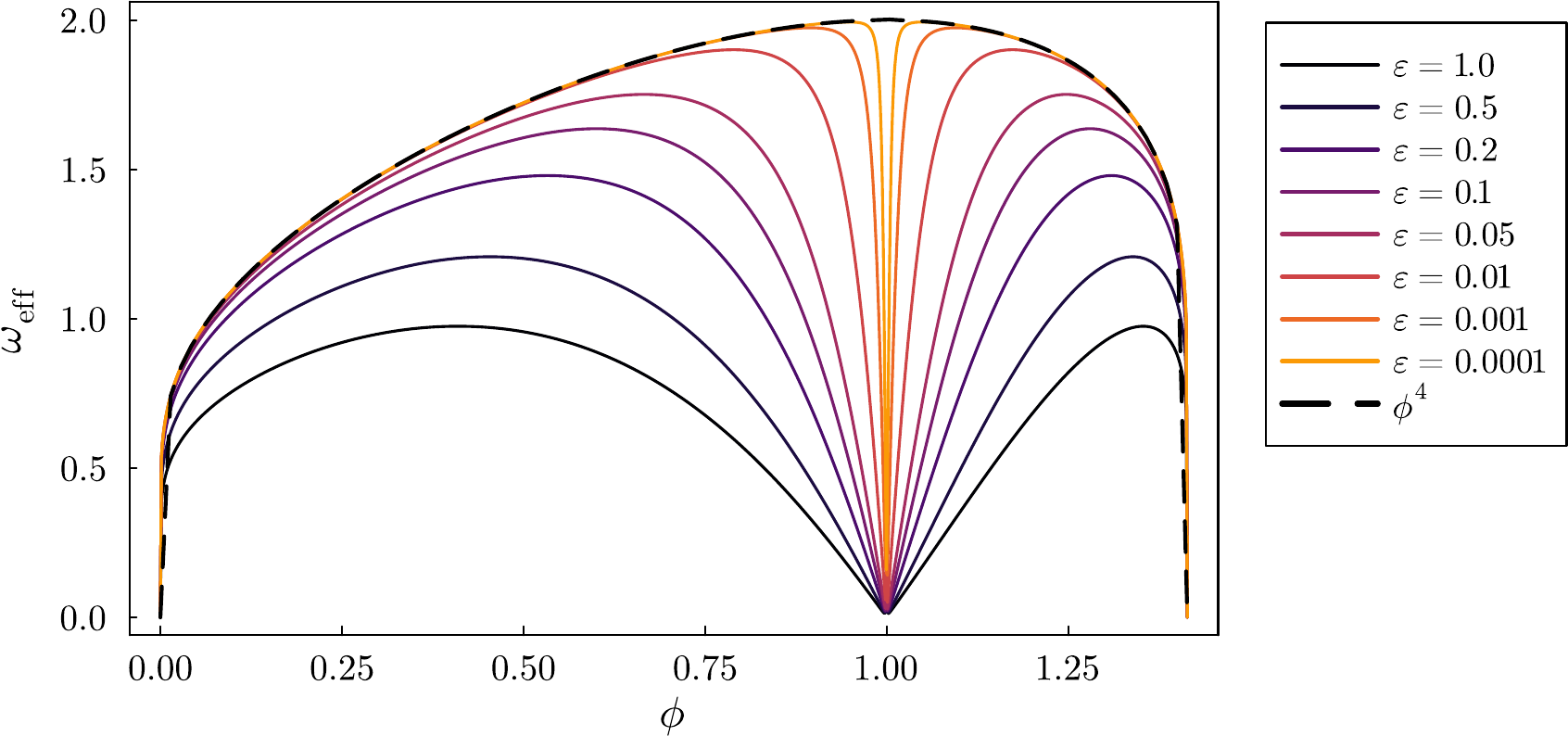}
 \caption{The effective frequency $\omega_\textrm{eff}$.} 
 \label{mass-plot-2}
 \vspace*{0.4cm}
  \includegraphics[width=1.0\columnwidth]{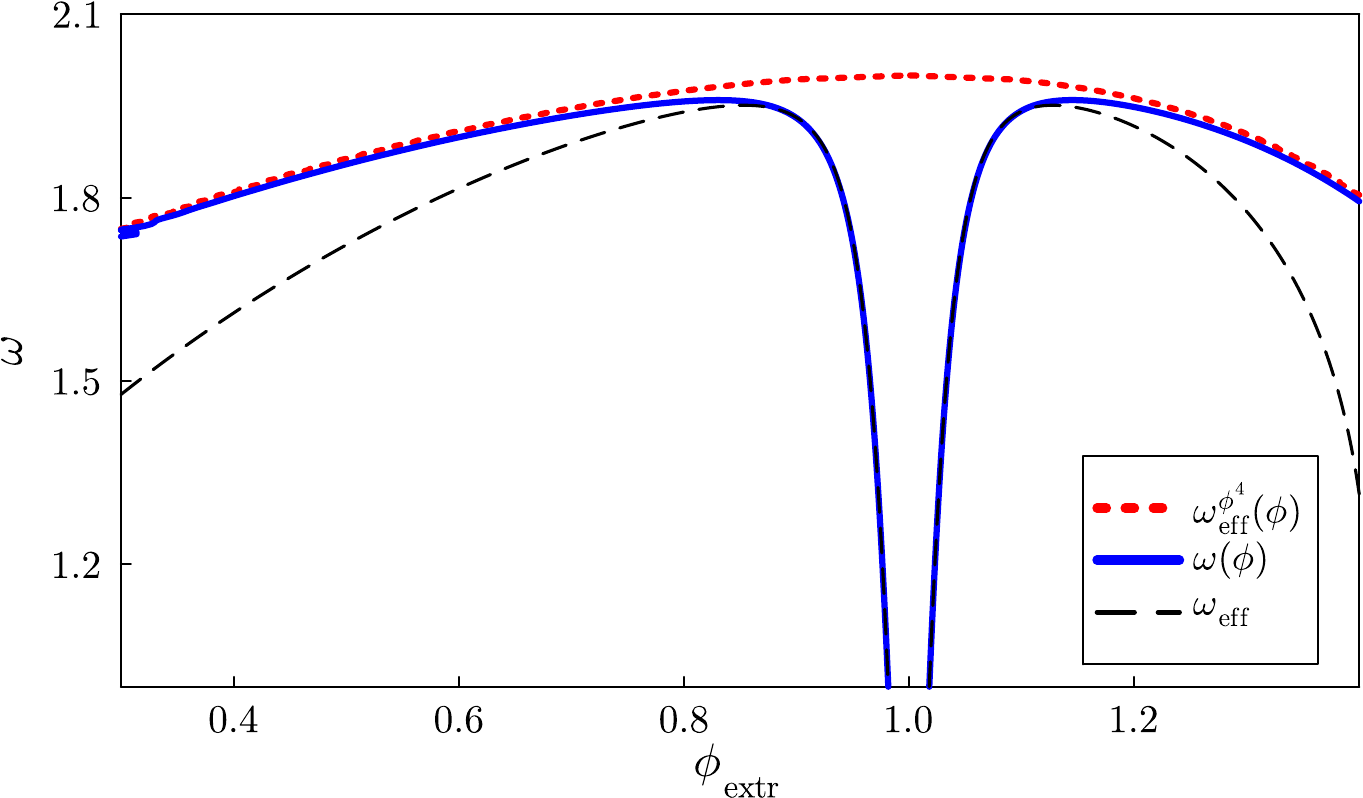}
 \caption{The dependence of measured oscillation frequency on the amplitude of the oscillon in the massless model with $\epsilon=0.003$ (blue curve) and in the $\phi^4$ theory. } 
 \label{mass-plot-comp}
 \end{figure}

The actual frequency of the oscillation may be also quite well reproduced using periodic motion approximation. We begin defining an effective period 
\begin{equation}
    T_\textrm{eff}(\phi) = 2\int_{\phi_1}^{\phi_2} \frac{d\phi}{\sqrt{2(U(\phi_1) - U(\phi))}}
\end{equation}
where $\phi_{1,2}$ are return points (i.e., amplitudes of the oscillon), which are related via a simple quadratic equation $\phi_1^2+\phi_2^2=2$. Then, the frequency is $\omega_\textrm{eff}= 2\pi / T_\textrm{eff}$. In Fig.\ \ref{mass-plot-2} we plot such frequencies for different values of $\epsilon$. They tend to zero value at the limit of infinitesimally small perturbations which obviously agrees with vanishing of the smeared (and perturbative) mass threshold. One can show that the leading behaviour is $\omega_\textrm{eff}=\Gamma(3/4)/(\Gamma(5/4)\sqrt{2\pi \epsilon })|\phi-1|\approx0.5394 |\phi-1| \epsilon^{1/2}$. Bigger perturbations correspond to nonzero values of $\omega_\textrm{eff}$. For too big amplitudes $\omega_\textrm{eff}$ again tends to zero value reflecting breaking of the usefulness of this quantity.

In Fig.\ \ref{mass-plot-comp}, we plot the actual frequency of oscillation of an oscillon found in the massless model with $\epsilon=0.003$ (blue curve) as a function of the oscillon amplitude i.e., the returning point $\phi_\textrm{extr}$. We compare it with a curve relating the frequency of oscillon with value of the returning point in the usual $\phi^4$ model (red dots) and with the effective frequency (dashed curve). Of course, during the time evolution the amplitude changes and we can measure how the actual frequency changes with the amplitude of the oscillon. We observe that the measured frequency reveals a transition from the frequencies identical as in the $\phi^4$ model (occurring for high amplitudes) to frequencies that agrees with $\omega_\textrm{eff}$ (for smaller amplitudes). As the amplitude tends to 0 we recover the massless regime. 

Finally, the last phase of the decay of the oscillon, where the frequency of the oscillations decreases after taking its critical value, reveals a self-similar property, see e.g., Fig.\  \ref{amplitud-plot}, lower panel. We found that the frequency of the oscillations decreases according to a universal law
\begin{equation}
    \omega(t)\sim t ^{-\beta}\,,\qquad \beta=0.37855\pm0.0006.
\end{equation}

\section{\label{sec:3dim}Three dimensional oscillons}
 It is crucial that the existence of oscillons in massless models is not exclusively bounded to (1+1) dimensions. On the contrary, they easily show up in higher dimensions as well. Furthermore, their appearance is due to exactly the same mechanism as in the lower-dimensional case. 
 
 Let us therefore consider our model (\ref{model}) in (3+1) dimensions. For simplicity we will assume spherically symmetric initial data, which results in a reduction of the dynamical problem to the radially symmetric one i.e., $\phi=\phi(r,t)$ and $(\nabla \phi)^2 = \phi_r^2$.  

\begin{figure}
\includegraphics[width=1.0\columnwidth]{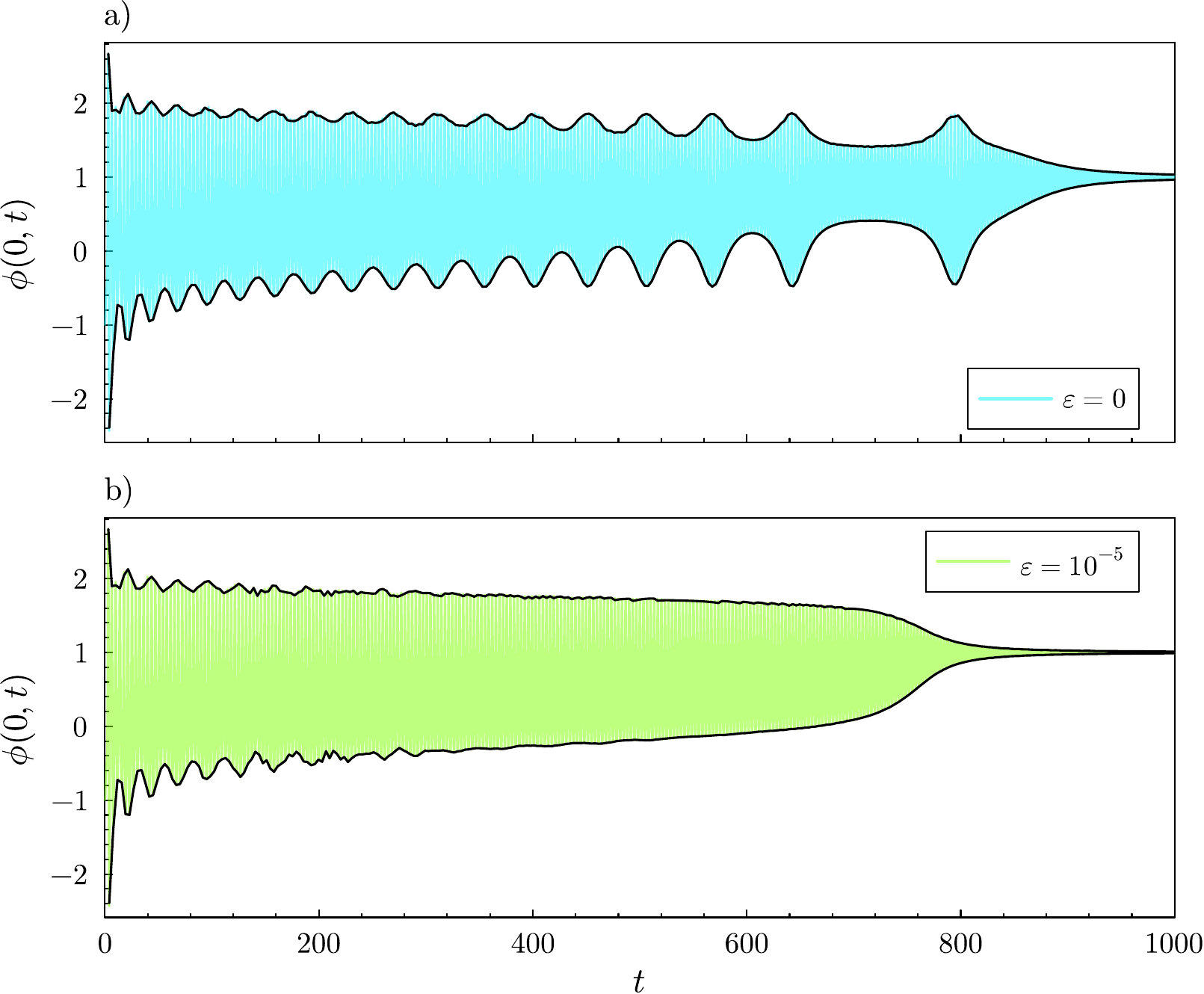}
 \caption{Decay of the amplitude of oscillons in (3+1) dimensions. {\it Upper:} the $\phi^4$ model i.e., $\epsilon=0$; {\it Lower:} the massless model (\ref{model}) with $\epsilon=10^{-5}.$} 
 \label{osc-3D}
 \end{figure}

 In Fig.\ \ref{osc-3D} we present time evolution of the field for radially symmetric initial data $\phi(r,0) = 1-2e^{-r^2/r_0^2}$ and $\partial_t \phi(r,0)=0$, with $r_0=2.849605$ which corresponds to the values studied in \cite{H}. The upper panel shows the evolution in the $\phi^4$ model, $\epsilon=0$, while in the lower panel we plot the evolution in the model with $\epsilon=10^{-5}.$ We see a large amplitude oscillon for the massless model, whose lifetime is comparable with the lifetime of the $\phi^4$ oscillon. 
 
\begin{figure}
  \includegraphics[width=1.0\columnwidth]{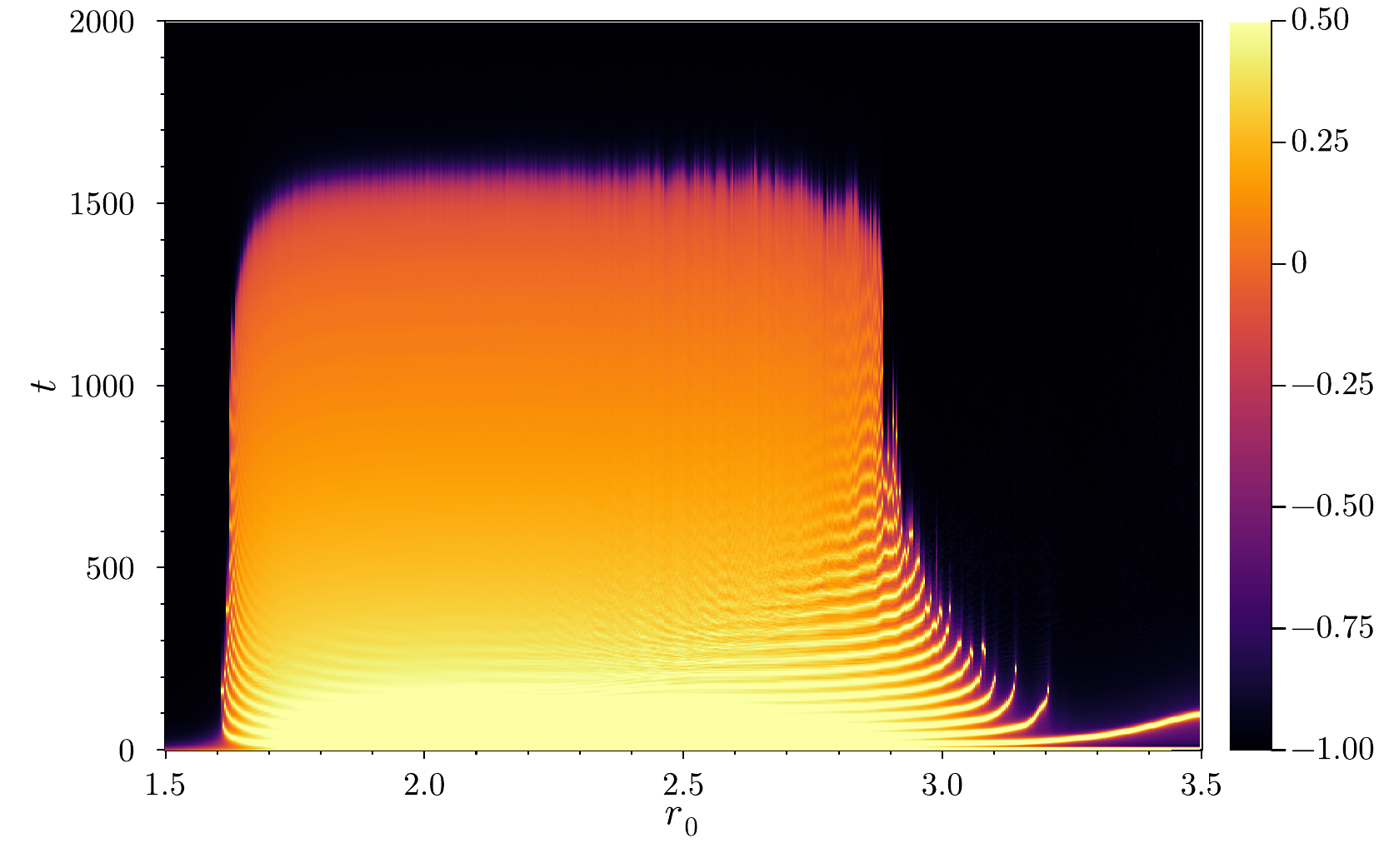}
\includegraphics[width=1.0\columnwidth]{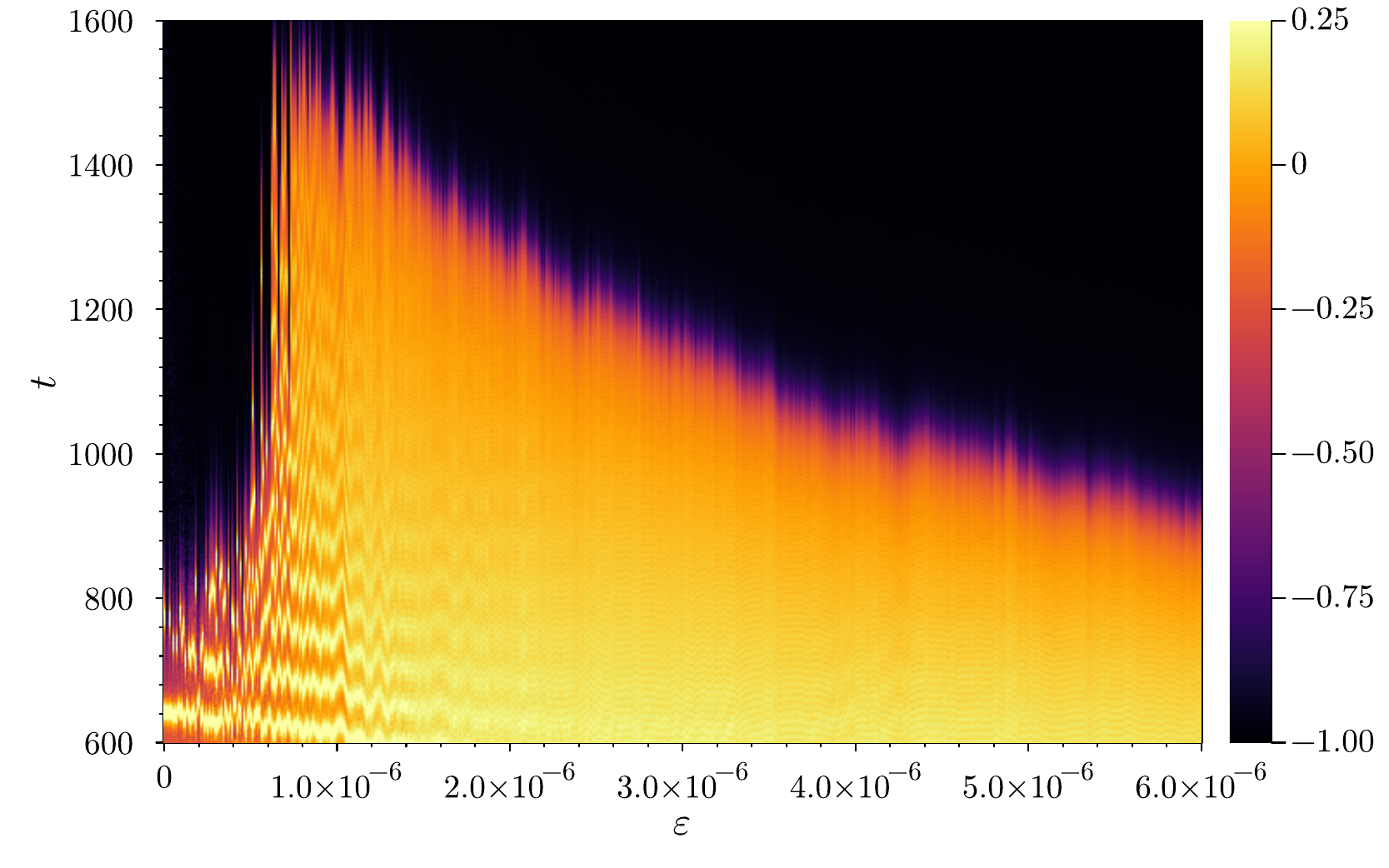}
 \caption{Dynamics and lifetime of the oscillon - envelope of $\phi(0,t)$. {\it Upper:} the massless model (\ref{model}) with $\epsilon=10^{-6}$ and different intial data parameters $r_0$. {\it Lower:} dynamics in different massless models and with $r_0=2.8496049$.} 
 \label{life}
 \end{figure}

 It is also visible that, in contrast to the $\phi^4$ case, this oscillon does not possess amplitude modulations in its late-time evolution. Indeed, initial modulations are smoothed out at $t \approx 200$ and from this moment we find very regular oscillations whose amplitude decreases very slowly until the oscillon enters into a rapid decay phase. This is a generic behaviour, existing for different initial perturbations, see Fig.\ \ref{life}, upper panel. It is known that the existence of such modulations is closely related with the lifetime of the oscillon and developing larger and larger modulations is the usual mechanism by which an oscillon is destabilized, see e.g., \cite{H} and \cite{GGB}. In the $\phi^4$ theory it has been shown that their absence, for example due to a choice of particular initial data, results in a slower decay of the oscillon \cite{H}. In the massless models (\ref{model}) we see the same phenomenon: smaller modulations lead to longer lifetimes. 
 
 In the lower panel of Fig.\ \ref{life} we show the lifetime of the oscillon for the initial data with $r_0=2.8496049$ but for $\epsilon \leq 6\cdot10^{-6}$. We see that the lifetime can grow significantly if we slightly change $\epsilon$. The resulting value may significantly exceed the lifetime in the $\phi^4$ theory. Hence, the oscillons not only do exist in massless models but can even be more stable. We also remark that in the small $\epsilon$ regime a resonant structure is visible. 
 
\section{\label{sec:summary}Summary}
In this work we have shown that oscillons can exist in massless field theories. This is an unexpected result which significantly enlarges the range of possible applications of oscillons. In particular we expect to find them in collisions of solitons with power-like decaying tails, see e.g. the so-called \emph{fat tail} kinks \cite{Kev, M-long}. 

Although the usual {\it perturbative} mass threshold is zero, suggesting nonexistence of oscillons, their appearance may be associated with a non-zero value of another \emph{effective} mass threshold, which takes into account finitely-sized perturbations. This quantity better describes the dynamical features of the model beyond the regime of infinitesimally small perturbations. It also affirms the nonlinear character of oscillons, for which the full nonlinearity of the field theoretical potential is more important than its near vacuum behaviour. 

It is challenging to understand these oscillons in the adiabatic invariant framework (I-balls) as  adiabaticity requires the scalar potential to be dominated by a quadratic term \cite{K}. This condition is clearly not satisfied in massless models where the quadratic term is absent. 

Finally, we underline that our results are relevant in any dimension. In fact, an oscilllon in a massless three dimensional model may have a significantly longer lifetime than one in the original massive theory. All of this opens new directions for the study of oscillons and their applications in the physics of nonlinear systems.
\section*{Acknowledgements}

TR and AW were supported by the Polish National Science Center, 
grant NCN 2019/35/B/ST2/00059 and the Priority Research Area under the
program Excellence Initiative—Research University at
the Jagiellonian University in Kraków. YS acknowledges support from the Hanse-Wissenschaftskolleg (Institute for Advanced Study) in Delmenhorst. PED was supported in part by the STFC under consolidated grant ST/T000708/1 “Particles, Fields and Spacetime”.

\section*{Supplementary material}

\subsection*{A. Lifetime}
\begin{figure}
  \includegraphics[width=1.0\columnwidth]{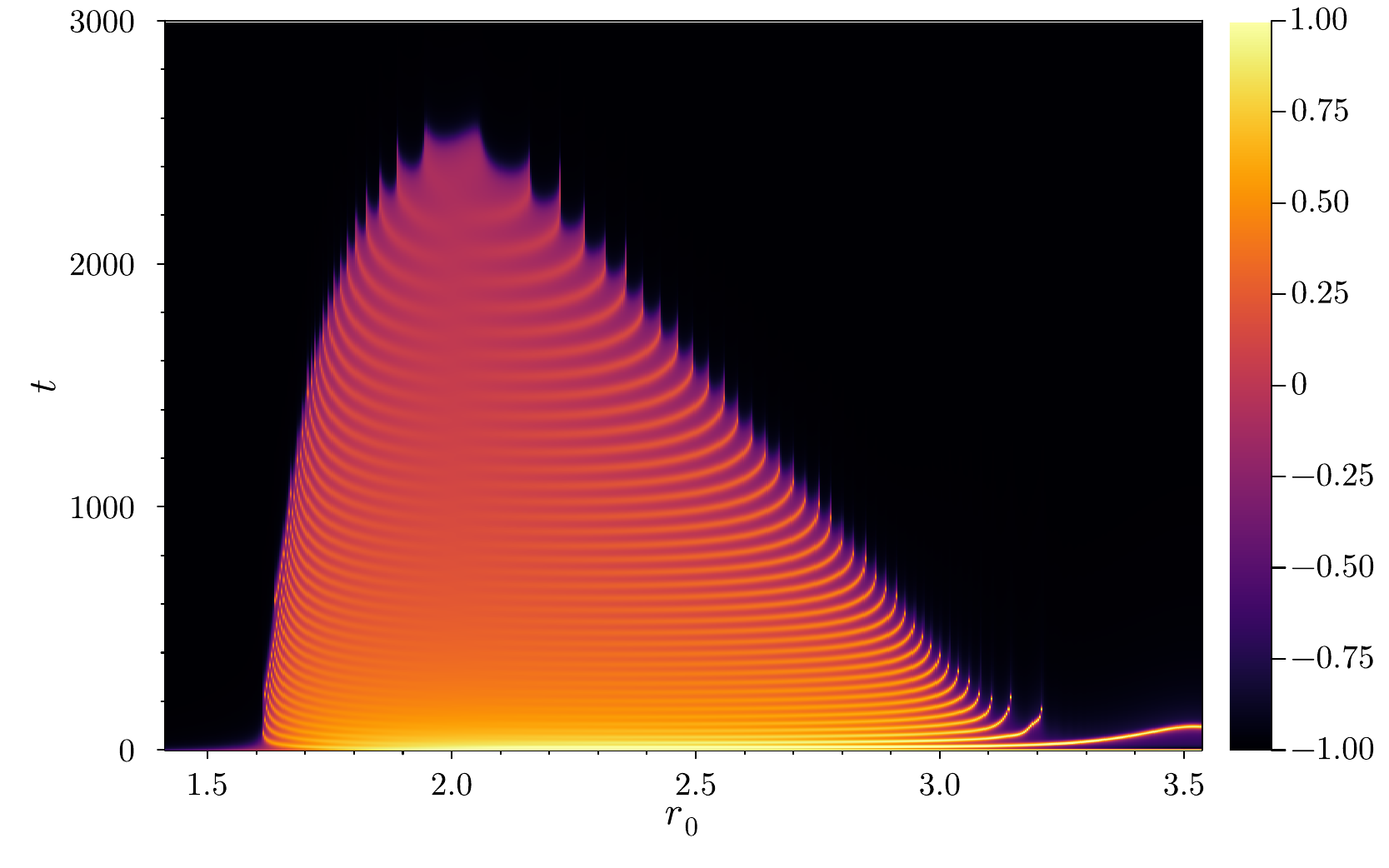}
 \caption{Dynamics and lifetime of the oscillon in $\phi^4$ in (3+1) dimensions - envelope of $\phi(0,t)$.} 
 \label{life-H}

\bigskip

  \includegraphics[width=1.0\columnwidth]{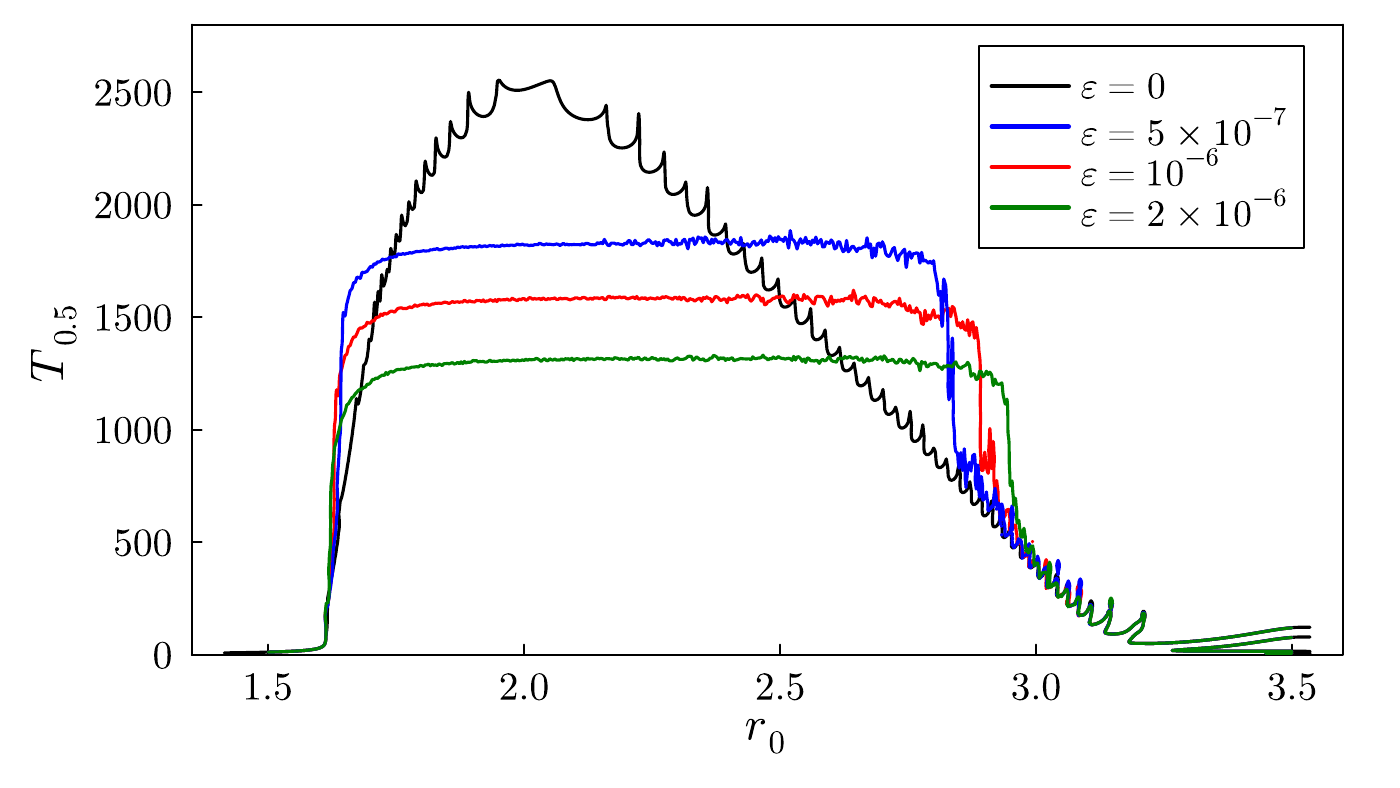}
 \caption{Comparison of the lifetime of the oscillon in (3+1) dimensions in $\phi^4$ theory and massless deformed models (\ref{model}). } 
 \label{life-comp}

 \bigskip
 
  \includegraphics[width=1.0\columnwidth]{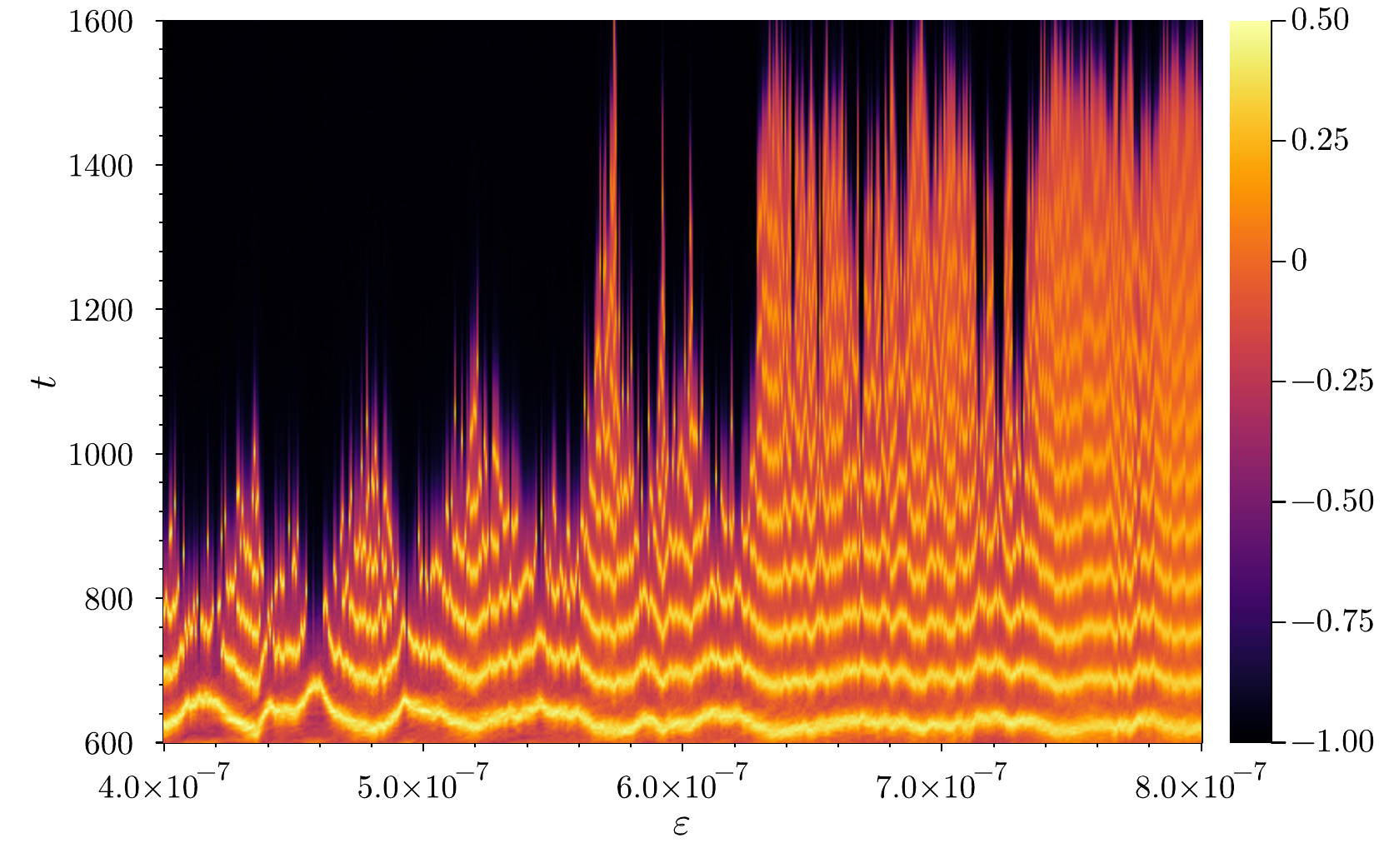}
 \caption{Envelope of $\phi(0,t)$ for different values of $\epsilon$ in $3+1d$ models for Gaussian initial conditions with $r_0= 2.8496049$.} 
 \label{life-res}
 \end{figure}
It is instructive to compare the lifetime of oscillons in the massless deformed model (\ref{model}) with the original $\phi^4$ theory in (3+1) dimensions. The Gaussian initial configuration is identical to that in \cite{H} and reads $\phi(r,0)=1-2e^{-2r^2/r_0^2}$, with $r_0$ being a positive parameter, $1.5<r_0<3.5$. The results for the $\phi^4$ theory are presented in Fig.\ \ref{life-H} and agree with  previous computations \cite{H}. If compared with the massless deformed model for $\epsilon=10^{-6}$, Fig.\ \ref{life}, upper panel, we see that both in the region of small $r_0$ and of large $r_0$ the massless deformation gives oscillons with longer lifetimes. This is also shown in Fig.\ \ref{life-comp}, where we compare the lifetimes of oscillons in the $\phi^4$ theory and in the massless deformed models with $\epsilon=5\cdot 10^{-7}$, $10^{-6}$ and $2 \cdot 10^{-6}$. 

In Fig.\ \ref{life-res} we present a zoomed version of the lifetime plot in the region with $\epsilon \leq 8\cdot 10^{-7}$, for $r_0= 2.8496049$. This clearly shows the appearance of a resonant effect. The lifetime of the oscillon depends chaotically  on the $\epsilon$ parameter and even an extremely small change in its value can have a very large impact on the oscillon decay.

 \begin{figure}
  \includegraphics[width=1.0\columnwidth]{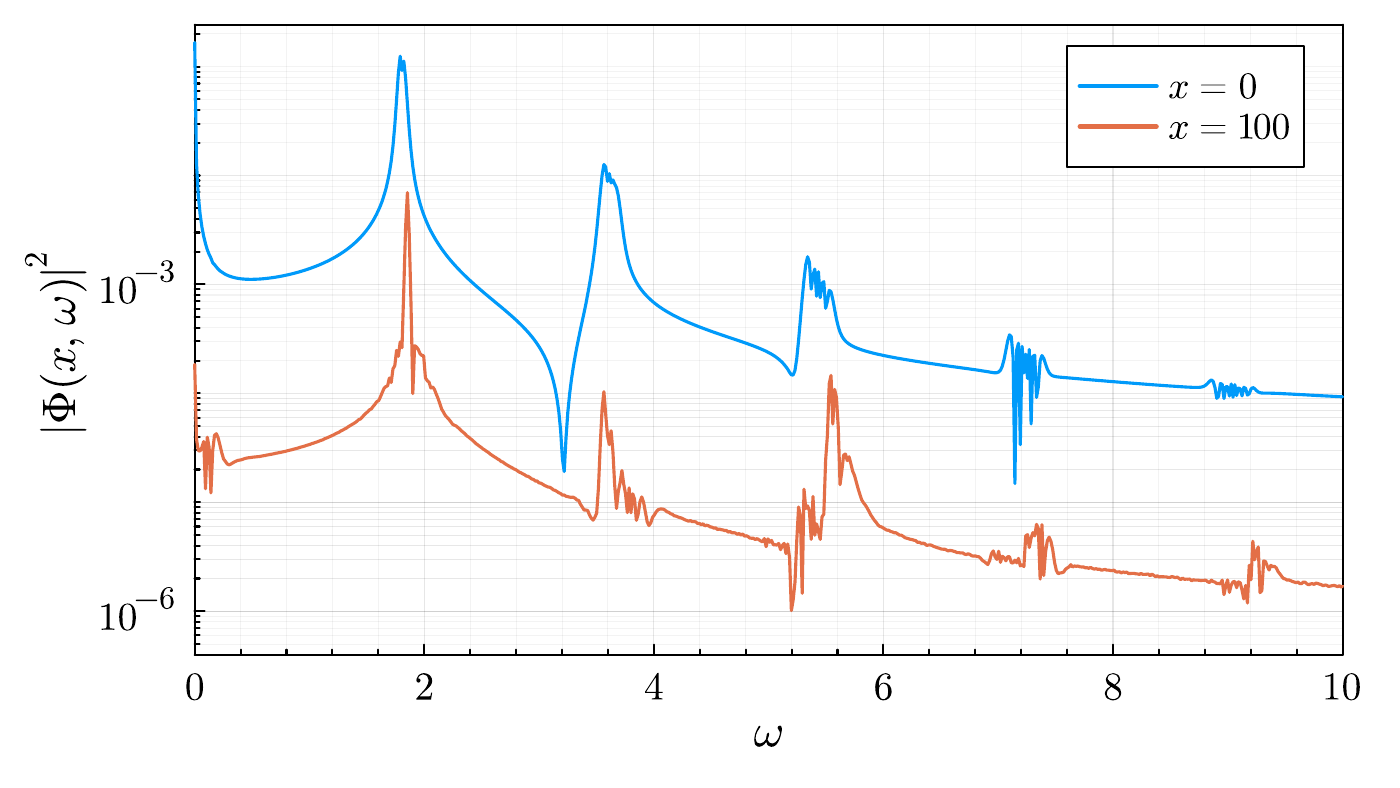}
 \caption{Power spectrum of an oscillon in (1+1) dimensions in the massless deformed model with $\epsilon=0.001$. } 
 \label{rad}
 \end{figure} 

\subsection*{B. Radiation from the oscillon}
We also checked that the oscillons in the massless deformed models (\ref{model}) lose their energy mainly via radiation through the fundamental frequency (first harmonics). This is clearly seen in Fig.\ \ref{rad} where we plot the power spectrum for the (1+1) dimensional oscillon for $\epsilon=0.001$. The power spectrum is computed at the origin (center of the oscillon) and in the far distance regime, $x=100$ for $t \in [200,600]$. In both cases, there is a peak at the fundamental frequency verifying that this is the main channel of dissipation of energy from the oscillon. 

\subsection*{C. Effective mass threshold}

It is important to underline that the effective mass threshold is only weakly dependent on a particular choice of the weight function $w(\phi)$. In Fig.\ \ref{rad} we compare the step function smearing with a Gaussian one. The observed differences in the values of $m_{\textrm{eff}}$ are very small, even in the large amplitude regime, and do not have any impact on presented analysis of the model. 

\subsection*{D. Movies}

\begin{itemize}
\item \href{arXiv link here}{\texttt{oscillon-extrema.mp4}} - evolution of field profiles at local maxima and minima of $\phi(0,t)$ for the 1+1 dimensional model with $\epsilon=0.003$, showing a relation between the radiation tails and the oscillon core.
\item \href{arXiv link here}{\texttt{lifetimes.mp4}} - evolution of the field at the centre, $\phi(0,t)$, for various values of $\epsilon$ in 3+1 dimensional models showing somewhat chaotic behaviour for small values of $\epsilon$  and decreasing modulations leading to increasing lifetimes. 
\end{itemize}
\newpage 

 \begin{figure}
  \includegraphics[width=1.0\columnwidth]{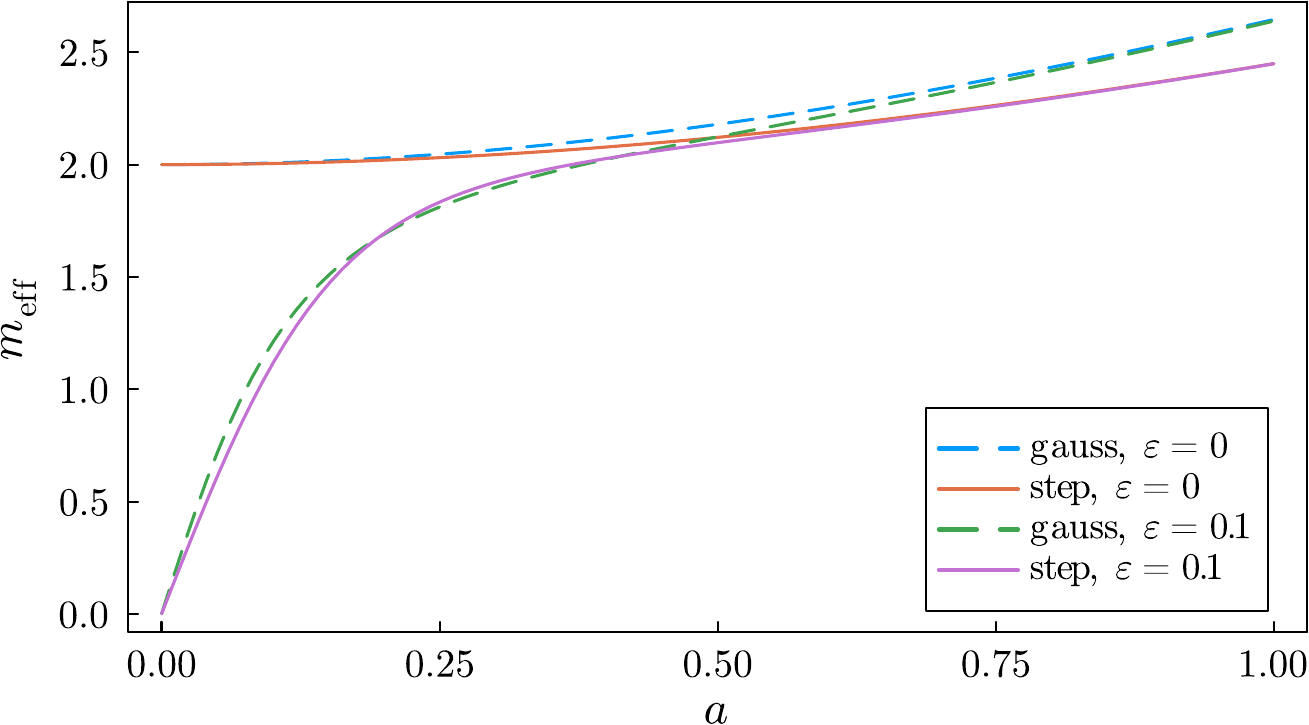}
 \caption{The effective mass threshold $m_{\textrm{eff}}$ for the step function and Gaussian smearing functions for the $\phi^4$ theory ($\epsilon=0$) and for the massless deformation with $\epsilon=0.001$. } \label{rad}
 \end{figure}

\end{document}